%% file: prl_wh1invfb.tex
\documentclass[aps,prl,preprint,superscriptaddress,showpacs]{revtex4}

\usepackage{graphicx}

\usepackage{ifpdf}
\ifpdf
 \DeclareGraphicsExtensions{.jpg,.pdf,.jpeg}
\else
\DeclareGraphicsExtensions{.eps,.ps,.eps.gz,.ps.gz}
\fi

\newcommand{\met}{\ensuremath{\not \!\! E_T}}

\begin{document}

\preprint{FERMILAB-PUB-07-567-E}

\title{Search for Standard Model Higgs Bosons Produced in Association with $W$ Bosons}

\input{July_2007_Authors_Visitors}


\date{\today}

\begin{abstract}
We report on the results of a search for standard model Higgs bosons
produced in association with $W$ bosons from $p\bar p$ collisions at
$\sqrt{s}=1.96\,\mathrm{TeV}$.  The search uses a data sample
corresponding to approximately $1\,\mathrm{fb}^{-1}$ of integrated
luminosity.  Events consistent with the $W\rightarrow \ell\nu$ and
$H\rightarrow b\bar b$ signature are selected by triggering on a
high-$p_T$ electron or muon candidate and tagging one or two of the
jet candidates as having originated from $b$ quarks.  A neural network
filter rejects a fraction of tagged charm and light flavor jets,
increasing the $b$-jet purity in the sample and thereby reducing the
background to Higgs boson production.  We observe no excess $\ell\nu
b\bar b$ production beyond the background expectation, and we set 95\%
confidence level upper limits on the production cross section times
branching fraction $\sigma(p\bar p\rightarrow WH) \cdot
Br(H\rightarrow b\bar b)$ ranging from 3.9 to 1.3 pb, for specific
Higgs boson mass hypotheses in the range 110 to
$150\,\mathrm{GeV}/c^2$, respectively.
\end{abstract}


\maketitle


The standard model of elementary particle physics (SM) provides for
electroweak gauge symmetry breaking via the Higgs
mechanism~\cite{Higgs:1964pj}, and the model predicts a single
physical remnant of the added Higgs field.  This remnant, the Higgs
boson $H$, has yet to be observed experimentally.  Results from direct
searches at the LEP collider exclude mass values less than
$114.4\,\mathrm{GeV}/c^2$ at a 95\% confidence
level~\cite{Barate:2003sz}, and global fits to precision electroweak
data exclude masses greater than $144\,\mathrm{GeV}/c^2$ at 95\%
confidence level~\cite{Alcaraz:2006mx}.  (Some models beyond the SM
predict Higgs bosons whose masses are not constrained by these
limits.)  For Higgs boson masses just above the range excluded by LEP,
the decay to bottom quarks $b\bar b$ dominates.  Even though gluon
fusion $gg
\rightarrow H
\rightarrow b\bar b$ has the largest cross section among Higgs
production processes in $p\bar p$ collisions~\cite{Han:1991ia}), the
$b\bar b$ data sample is dominated by non-resonant multi-jet
background.  Consequently, we search for $WH$ production, requiring a
leptonic $W$ boson decay to suppress the background.  In this Letter
we report results of a search for low-mass SM Higgs bosons produced in
association with $W$ bosons and decaying to $b\bar b$ pairs.  The
resulting $\ell
\nu b\bar b$ final state is identified by selecting events with
exactly one high-energy electron or muon candidate, large missing transverse
energy, and one or two jet candidates having a secondary vertex characteristic
of heavy quark decay.

Recent searches at CDF and D0~\cite{Abulencia:2005ep,Jonsson:2006bg}
were limited not only by smaller data samples, but also by
contamination from jets associated with charm or light quarks which
are falsely tagged as $b$ jets.  The search described in this Letter
employs for the first time a neural network filter to reject such
events, thereby improving the purity of the selected event sample.
The data sample of $p\bar p$ collisions at
$\sqrt{s}=1.96\,\mathrm{TeV}$ used here corresponds to
$0.955\pm0.057\,\mathrm{fb}^{-1}$ of integrated luminosity, nearly
three times the sample used in previous searches.


The CDF II detector is a general-purpose detector located at the
Tevatron $p\bar p$ collider at
Fermilab~\cite{Acosta:2005am,Acosta:2004hw}.  It consists of a
cylindrical magnetic spectrometer surrounded by sampling calorimeters
used to measure energies of electromagnetic showers and jets.  Charged particle
tracking is performed with microstrip silicon detectors surrounded by a large
cylindrical multilayer drift chamber, both immersed in a solenoidal magnetic
field.  Jets are identified as a collection
of hadronic and electromagnetic calorimeter towers, which are clustered
using an iterative cone algorithm with a cone of $\Delta R =
\sqrt{(\Delta\phi)^2 + (\Delta\eta)^2} = 0.4$ 
units in the azimuth-pseudorapidity space~\cite{coordinates,
Abe:1991ui}.  Planar drift chambers used for
muon detection surround the calorimeters at
least five interaction lengths from the interaction region.


Events are collected using high-$p_T$ electron or muon triggers with a
three-level selection filter.  The first- and second-level criteria
ensure that purely electromagnetic calorimeter clusters exist or that
track stubs in the muon chambers align with drift chamber tracks
having transverse momentum at least $8\,\mathrm{GeV}/c$.  The
third-level trigger ensures that a fully-reconstructed track with
$p_T$ at least $18\,\mathrm{GeV}/c$ points to the electromagnetic
cluster or muon stub.


Events compatible with the $\ell \nu b \bar b$ final state are
selected by requiring exactly one electron or muon candidate and
missing transverse energy $\met>20\,\mathrm{GeV}$, after jets are corrected for detector imperfections and non-linear calorimeter response~\cite{coordinates}.
The electron or muon must be within the central part of the detector,
in the pseudorapidity regions $|\eta|<1.1$ or $|\eta|<1.0$,
respectively, and must have transverse energy greater than
$20\,\mathrm{GeV}$.  The lepton must be isolated from the rest of the
event by a cone of radius $\Delta R=0.4$ containing no more than 10\%
of the lepton energy (excluding the lepton itself).
It must also be no more than $5\,\mathrm{cm}$ in $z$ away from the
primary event vertex, which is defined by fitting a subset of charged
particle tracks in the event to a single vertex.  To suppress
background from $Z$ boson and diboson production, we reject events
with more than one isolated lepton, as well as events in which the
lepton and another high-energy track of opposite sign form an
invariant mass between 76 and $106\,\mathrm{GeV}/c^2$.  The
requirements for the lepton transverse energy and missing transverse
energy reject multi-jet background while maintaining efficiency for
the $WH$ signal.  Jets used in the analysis must fall within the
acceptance of the silicon detector ($|\eta|<2.0$) for reliable
$b$-tagging, and they must have transverse energy greater than
$15\,\mathrm{GeV}$.  Even though the $W+2$ jet final state is the
target sample for this search, other samples with $W$+1 or $W$+3, 4
jets are useful for cross-checks of the background estimates with
similar topologies.


A $B$ hadron, with relatively long lifetime and large mass, can decay
to charged particles whose tracks have large impact parameter, the
distance of closest approach to the interaction point in the
transverse plane.  Such tracks are fit to a secondary vertex, and the
decay length of the $B$ hadron is defined as the distance between this
vertex and the primary vertex.  Specifically, we apply the {\sc
secvtx} secondary vertex finding algorithm~\cite{Abulencia:2006in} to
each jet in the event, using tracks within the $\Delta R=0.4$ cone
centered on the jet axis.  Three tracks with impact parameter
significances greater than 2.0 are fit to a decay vertex.  If this
first pass fails, a second pass is attempted with two tracks having
impact parameter significances greater than 3.0.  To limit the number
of secondary vertices stemming from material interactions, we reject
vertices at radii greater than $2.5\,\mathrm{cm}$, as well as any
vertices reconstructed within the material of the beampipe and the
innermost silicon layer ($1.2\,\mathrm{cm} < r < 1.5\,\mathrm{cm}$).
Jets are $b$-tagged if the magnitude of the significance of the
transverse decay length is greater than 7.5.  Jets with a negative
decay length have a reconstructed flight direction opposite the jet
direction.  This can happen when tracks coming from the primary vertex
have significantly mismeasured impact parameters.

In addition to the secondary vertex finding algorithm, a neural
network (NN) filter has been trained with the {\sc jetnet}
program~\cite{Peterson:1993nk} to reject tagged jets originating from
charm or light ($u,d,s$) quarks.  When a $b$ quark hadronizes,
particles originating at the $B$ hadron vertex tend to carry a large
fraction of the jet energy.  They also tend to have a higher vertex
mass -- the invariant mass of all tracks in the secondary vertex --
and longer decay lengths than in jets from light quarks.  The NN
filter uses these jet characteristics to discriminate between $b$ jets
and $c$ or light flavor jets.  It is composed of two networks in
series, one to separate $b$ jets from light quark jets, and the second
to separate $b$ jets from $c$ jets.  Both networks have the same set
of 16 inputs: the number of tracks in the secondary vertex, the
$\chi^2$ value of the vertex fit, the transverse decay length and its
significance, the vertex mass calculated by assuming the charged pion
mass for all particles, the proper time assuming the vertex mass, the
fraction of the jet $p_T$ carried by tracks in the vertex, the vertex
pass number, the number of tracks with significant impact parameter,
the reconstructed mass of the {\sc secvtx} pass 1 and pass 2 tracks,
the numbers of pass 1 and pass 2 tracks, the fraction of the jet $p_T$
carried by the pass 1 and pass 2 tracks, and finally the probability
of a selected ensemble of tracks to have originated at the primary
vertex~\cite{Abulencia:2006kv}.  The selection cuts on the NN output
are chosen to give 90\% efficiency for true $b$ jets identified with
the secondary vertexing algorithm.  The corresponding rejection
factors are $65\pm5\%$ for light flavor jets and $50\pm5\%$ for charm
jets, as measured using simulated events and verified with multijet
data.


Our search criteria select events with exactly one high-energy charged
lepton, missing transverse energy, and two jets.  The search
sensitivity is maximized by defining two distinct subsamples based on
the following $b$-tagging requirements: single-tagged events with
exactly one $b$-tagged jet which passes the NN filter, and
double-tagged events with two $b$-tagged jets.  Because events with
charm and light-flavor jets are unlikely to be double-tagged, the
extra NN filter is not applied to double-tagged events.  The selected
event sample includes contributions from other SM processes.  The
largest background rates are due to $W$+jets production, $t\bar t$
production, and non-$W$ multijet production, with small contributions
from electroweak boson production $WW$ or $WZ$.

The dominant background contribution comes from $W$+jets production,
either with jets from $b$ or $c$ quarks or with jets mistagged by the
$b$-tagging algorithm.  The effect of true $W$+heavy-flavor production
is estimated from a combination of data and simulation.  We use the
{\sc alpgen} Monte Carlo program~\cite{Mangano:2002ea} to calculate
the rate of $Wb\bar b, Wc\bar c$, and $Wc$ production relative to
inclusive $W$+jets production.  Then this relative rate is applied to the
observed $W$+jets sample, after non-$W$ and $t\bar t$ contributions
have been subtracted.  Finally, we apply a $b$-tagging efficiency and
NN filter rate derived using the {\sc alpgen} event samples.

Events from $t\bar t$ production followed by leptonic $W$ decay
typically have two $b$ jets from $t$ decay, significant missing
transverse energy, and one or two high-energy leptons with two or zero
additional jets, depending on whether one or both $W$ bosons from the
top quarks decay leptonically.  Our selected sample includes
contributions from pairs of top quarks which both decay leptonically,
but for which the second lepton is not reconstructed.  It also
includes contributions from pairs of top quarks with one leptonic
decay which are selected as two-jet events because two out of four
jets do not satisfy the selection criteria.  The contribution from
$t\bar t$ production to the $\ell\nu b \bar b$ final state is
estimated using simulated {\sc pythia} events~\cite{Sjostrand:2000wi}.
It is normalized to the NLO cross section
$6.7^{+0.7}_{-0.9}\,\mathrm{pb}$ calculated for
$m_t=175\,\mathrm{GeV}/c^2$~\cite{Cacciari:2003fi}.  The small
contribution from production of single top quarks is estimated using
{\sc madevent}~\cite{Maltoni:2002qb} and {\sc pythia} normalized to
the NLO cross section~\cite{Harris:2002md}.

Multijet events may have high-energy identified leptons or missing
transverse energies, both mimicking the signature of $W$ decay.  These
may be from semileptonic heavy flavor decay or from false
reconstructions.  The identified leptons from such events are rarely
isolated in energy, as required by our event selection, and seldom
yield large missing transverse energy.  We therefore calculate the
number of non-$W$ events in our selected sample by extrapolating from
sideband regions (defined in the space of lepton energy isolation and
missing transverse energy) into the signal
region~\cite{Acosta:2004hw}.

Contributions from events with falsely tagged light-flavor jets are
estimated by measuring a false tag (or mistag) rate in generic jet
data.  To first order the negative tag rate is a good approximation
of the mistag rate because light-flavor jets, whose tracks are
prompt, have Gaussian reconstructed decay length distributions
symmetric about zero~\cite{Abulencia:2006in}.  The mistag
rate is further modified by the NN filter efficiency.  The resulting
overall mistag rate is applied to the $W$+jets sample to yield the
number of mistagged events present in the sample.

Small contributions from electroweak backgrounds ($WW$, $ZZ$, $WZ$,
and $Z\rightarrow \tau \tau$) are estimated using the most recent
theoretical cross section calculations~\cite{Campbell:1999ah}, with
acceptances calculated using fully simulated events from the {\sc
pythia} Monte Carlo program.

The dominant uncertainty in the $W$ + heavy flavor background is the
calibration factor for simulation derived from multijet
data~\cite{Acosta:2004hw}.  Different simulation inputs give different
factors, and we find a 35\% relative error on the background from
heavy flavor.  The background from false tags has major uncertainties
on the rate correction due to particle interactions in detector
material and on the NN rejection factor.  Both are 15\% relative
errors.  Cross-checks of sideband data yield a 17\% relative
uncertainty on the non-$W$ multijet estimate.  The electroweak
background estimates for diboson and single top are subject to
uncertainties in the $b$-tagging efficiency and the cross section
predictions.

We use the large $b$-tagged sample of $W$+1 jet events to derive a
data-based scaling factor of $1.2\pm 0.2$, which corrects a residual
mismatch between the heavy flavor fraction correction factor in
multijet data and the $W$+jets sample.  This single factor is applied
to the $W$+ heavy flavor background calibration for all jet
multiplicities, and it improves the agreement for the sideband
multiplicities of $W$+1, 3, 4 jets.
A summary of the estimated background contributions to the lepton +
jets sample is shown in Table~\ref{tbl:Njets_snntag_0d0h0i_All}, along
with the results from the data sample.

\begin{table*}
\begin{center}
\begin{tabular}{cccccccc}
\hline \hline
                          Selection & \multicolumn{4}{c}{single NN tag} & \multicolumn{3}{c}{double tag} \\
                   Jet Multiplicity &                1 jet &                2 jets &               3 jets &       $\geq4$ jets &               2 jets  &              3 jets  &        $\geq4$ jets \\
\hline
                             Mistag &     139.7 $\pm$ 27.3 &      53.9  $\pm$ 10.7 &      15.7 $\pm$  3.1 &       4.2 $\pm$  0.8 &      3.5 $\pm$  0.5 &      2.0  $\pm$  0.3 &       1.2 $\pm$  0.2 \\
                        $Wb\bar{b}$ &     306.9 $\pm$ 106.9 &     144.7 $\pm$ 49.4 &      29.9 $\pm$  9.7 &       6.4 $\pm$  2.5 &     20.3 $\pm$  7.0 &      5.7  $\pm$  1.8 &       1.0 $\pm$  0.4 \\
                        $Wc\bar{c}$ &      63.1 $\pm$ 22.0 &      43.0  $\pm$ 14.7 &       8.7 $\pm$  2.8 &       1.9 $\pm$  0.8 &      3.3 $\pm$  1.1 &      0.4  $\pm$  0.1 &       0.1 $\pm$  0.04 \\
                               $Wc$ &     185.7 $\pm$ 47.2 &      34.4  $\pm$  9.0 &       3.4 $\pm$  0.9 &       0.6 $\pm$  0.2 &      0.0 $\pm$  0.0 &      0.0  $\pm$  0.0 &       0.0 $\pm$  0.0 \\
                  $t\bar{t}$(6.7pb) &       6.9 $\pm$  1.2 &      42.0  $\pm$  6.6 &      84.9 $\pm$ 12.8 &      98.6 $\pm$ 14.3 &     10.4 $\pm$  2.3 &      29.5 $\pm$  6.4 &      45.5 $\pm$  9.9 \\
                         Single Top &      16.7 $\pm$  1.8 &      23.5  $\pm$  2.4 &       4.8 $\pm$  0.5 &       0.8 $\pm$  0.1 &      4.2 $\pm$  0.7 &       1.4 $\pm$  0.2 &       0.3 $\pm$  0.1 \\
Diboson/$Z^{0}\rightarrow \tau\tau$ &      11.7 $\pm$  2.2 &      14.2  $\pm$  2.3 &       3.9 $\pm$  0.9 &       1.0 $\pm$  0.3 &      1.2 $\pm$  0.3 &       0.3 $\pm$  0.1 &       0.1 $\pm$  0.1 \\
                        non-$W$ QCD &      84.2 $\pm$ 14.1 &      38.9  $\pm$  6.7 &      12.1 $\pm$  2.3 &       5.5 $\pm$  1.2 &      1.4 $\pm$  0.3 &       0.9 $\pm$  0.2 &       0.3 $\pm$  0.1 \\
\hline
                   Total Background &     814.9 $\pm$ 140.7 &     394.4 $\pm$ 66.6 &     163.4 $\pm$ 18.7 &     118.9 $\pm$ 14.9 &  44.2 $\pm$  8.5 &      40.1 $\pm$  6.8 &      48.6 $\pm$  10.0 \\
\hline
        Observed Events &                    856 &                    421 &                    177 &                    139 &                 39 &                     44 &                     65 \\
\hline \hline
\end{tabular}
      \caption{Background estimate for events with exactly one NN
  $b$-tag or double tag as a function of jet multiplicity.
  \label{tbl:Njets_snntag_0d0h0i_All}
  } 
\end{center} 
\end{table*}

\begin{figure}[htbp]
 \begin{center}
  \includegraphics[width=0.48\textwidth]{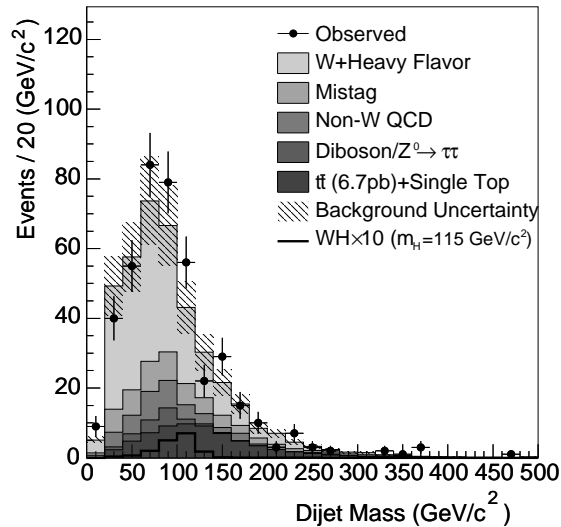}
  \caption{Reconstructed dijet mass distributions for $W$+2-jet events with a
  single $b$-tag passing the NN filter.  The histogram
  binning is the same used in the binned likelihood calculation.}
  \label{fig:1tagmass} \end{center}
\end{figure}

\begin{figure}[htbp]
 \begin{center}
  \includegraphics[width=0.48\textwidth]{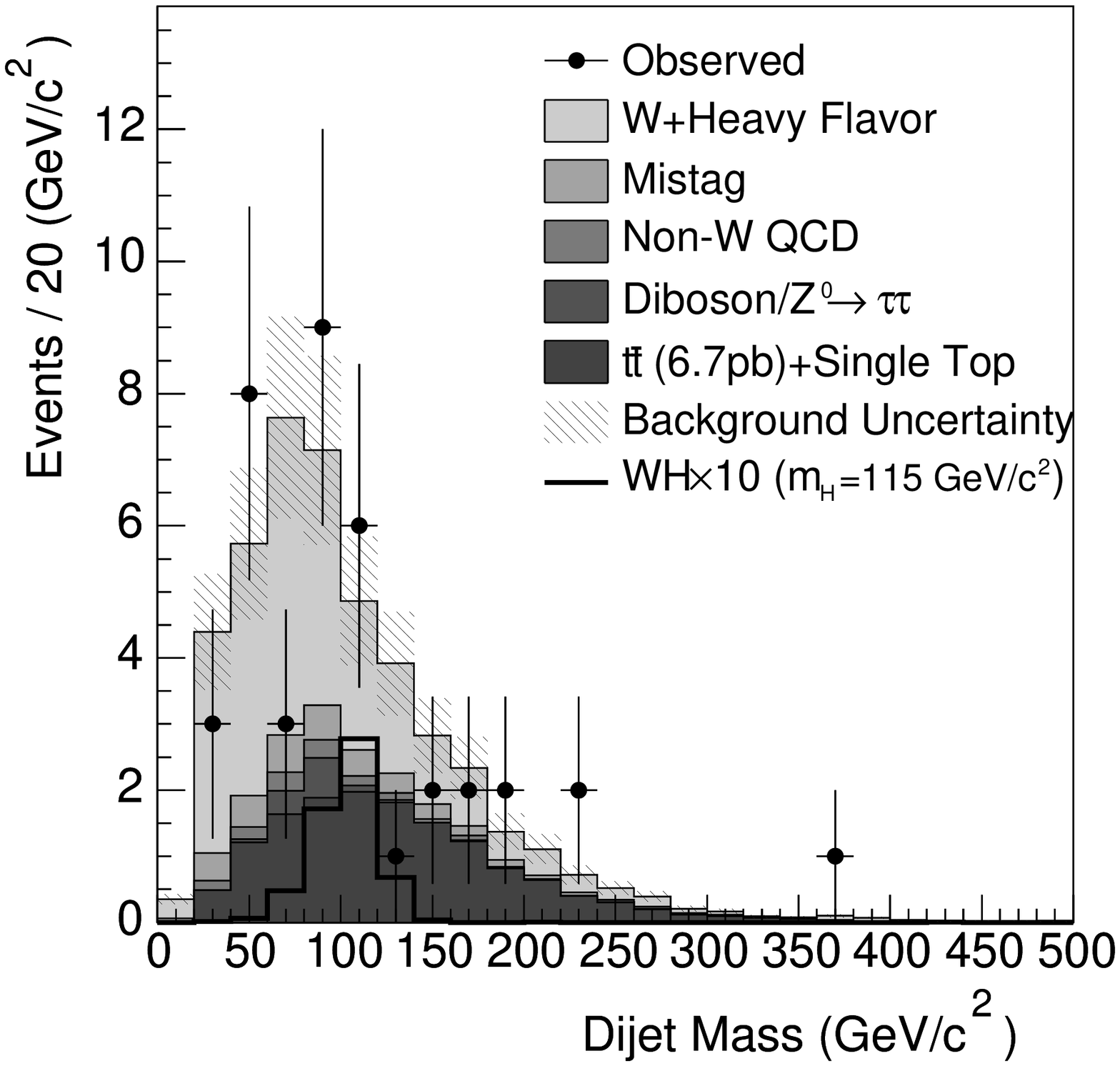}
  \caption{Reconstructed dijet mass distributions for $W$+2-jet events with two
  $b$-tagged jets.  The histogram binning is the same used in the
  binned likelihood calculation.}  \label{fig:2tagmass} \end{center}
\end{figure}

The signal process in which a Higgs boson decays to $b\bar b$ is
expected to show a resonant peak in the dijet mass spectrum.
Figures~\ref{fig:1tagmass} and~\ref{fig:2tagmass} show the dijet mass
spectra in the single- and double-tagged 2-jet samples for the estimated
background as well as for the observed events.  A
$115\,\mathrm{GeV}/c^2$ Higgs boson signal at ten times the SM rate is
shown for comparison.  There is no significant excess observed in the
dijet mass spectrum.  The largest discrepancy, for masses near
$100\,\mathrm{GeV}/c^2$, is less than one standard deviation defined
by the uncertainty on the background estimate.

The acceptance for $WH \rightarrow \ell \nu b\bar b$, including
leptonic $\tau$ decays, is calculated from samples generated with the
{\sc pythia} Monte Carlo program using Higgs boson mass values between
$110$ and $150\,\mathrm{GeV}/c^2$.  Additional efficiency factors
include the trigger and identification efficiencies measured
separately for the lepton types, and a $b$-tagging efficiency
data-to-simulation scale factor.  The acceptances for the single NN
tag and double tag selections are $1.3\pm0.1\%$ and $0.4\pm0.1\%$,
including the $W$ branching ratio to lepton pairs, for a mass
hypothesis of $115\,\mathrm{GeV}/c^2$.  The dominant systematic
uncertainty on the acceptance is the $b$-tagging scale factor
uncertainty, which is a 5.3\% relative error for the single-tagged
selection and a 16\% relative error for the double-tagged selection.
Variations in the final state radiation model introduce relative
uncertainties of 3.2\% and 8.6\% into the NN single-tagged acceptance and
double-tagged acceptance, respectively.  Additional sources of systematic
error include the jet energy scale, the lepton identification
efficiency, and the initial state radiation
model~\cite{Abulencia:2006in,Bhatti:2005ai}.

Limits on the number of Higgs boson events, interpreted as the
production rate times the branching fraction, are derived using a
binned likelihood technique assuming Poisson statistics.  The sample
is divided into bins of reconstructed dijet mass because the Higgs
boson production signal is expected to show a mass resonance.  By
counting the number of events in each reconstructed mass bin, we
discriminate more effectively between the peaked signal and the
predicted background shape.  Likelihoods are calculated separately for
the single-tagged and double-tagged selections.  The best exclusion
comes from a combination of the single-tagged selection with the NN
filter applied and the double-tagged selection.

To calculate the production limits, a Bayesian interval is constructed
from the cumulative likelihood distributions and a prior probability
density function uniform in the number of Higgs boson signal events
$s$.  The 95\% confidence level upper limit is defined to be the value
$s_\mathrm{up}$ for which $\int_0^{s_\mathrm{up}} L(s) ds /
\int_0^\infty L(s) ds = 0.95$.  
The number of signal events is then converted to a Higgs boson
production cross section times branching fraction $\sigma(p\bar
p\rightarrow WH) \cdot Br(H\rightarrow b\bar b)$.

\begin{figure}[htbp]
 \begin{center}
  \includegraphics[width=0.48\textwidth]{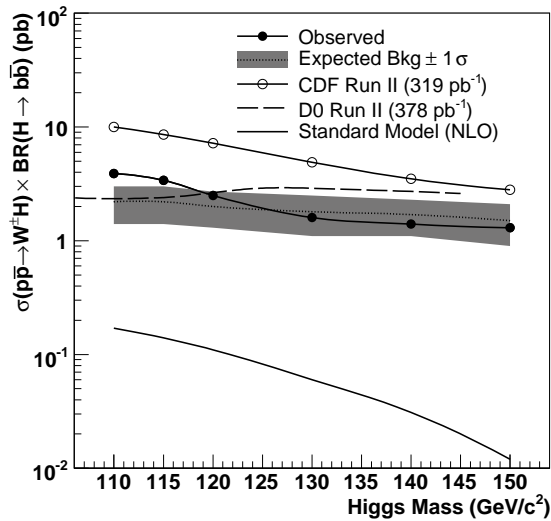}
  \caption{95\% confidence level upper limit on Higgs boson production
  cross section times branching fraction as a function of Higgs boson
  mass hypothesis.  The expected limits from background-only
  pseudoexperiments are shown in addition to the observed results from
  this search and previous CDF and D0 searches.}  \label{fig:limit}
  \end{center}
\end{figure}

The observed 95\% confidence level upper limits on the cross section
times branching fraction range from 3.9 to $1.3\,\mathrm{pb}$, for
Higgs boson mass hypotheses from 110 to $150\,\mathrm{GeV}/c^2$,
respectively.  Figure~\ref{fig:limit} summarizes the observed limits
as well as the expected limits as a function of the Higgs boson mass
hypothesis.  A set of background-only pseudoexperiments is generated
for each mass; the median limit value in this set defines the expected
limit in the absence of signal, and the spread of the
pseudoexperiments defines the $1\sigma$ band.  The observed limit in
the low mass region is roughly 2 standard deviations higher than the
expected limit.


In this Higgs boson search, we have employed a novel neural network
$b$-tagging filter on a dataset nearly three times the size of
previous searches.  The resulting exclusion improves significantly the
limits on the allowed production rate for Higgs bosons in $p\bar p$
collisions.  Even though the largest improvement by far comes from the
larger dataset, separating the single- and double-tag samples results
in a 20\% improvement beyond the previous analysis, and rejecting
charm and light-flavor jets with the NN gains another 5\% in
sensitivity.

\begin{acknowledgments}
  We thank the Fermilab staff and the technical staffs of the
  participating institutions for their vital contributions.  This work
  was supported by the U.S. Department of Energy and National Science
  Foundation; the Italian Istituto Nazionale di Fisica Nucleare; the
  Ministry of Education, Culture, Sports, Science and Technology of
  Japan; the Natural Sciences and Engineering Research Council of
  Canada; the National Science Council of the Republic of China; the
  Swiss National Science Foundation; the A.P. Sloan Foundation; the
  Bundesministerium f\"ur Bildung und Forschung, Germany; the Korean
  Science and Engineering Foundation and the Korean Research
  Foundation; the Science and Technology Facilities Council and the
  Royal Society, UK; the Institut National de Physique Nucleaire et
  Physique des Particules/CNRS; the Russian Foundation for Basic
  Research; the Comisi\'on Interministerial de Ciencia y
  Tecnolog\'{\i}a, Spain; the European Community's Human Potential
  Programme; the Slovak R\&D Agency; and the Academy of Finland.
\end{acknowledgments}


\input{prl_wh1invfb.bbl}
\end{document}

%% file: July_2007_Authors_Visitors.tex
\affiliation{Institute of Physics, Academia Sinica, Taipei, Taiwan 11529, Republic of China} 
\affiliation{Argonne National Laboratory, Argonne, Illinois 60439} 
\affiliation{Institut de Fisica d'Altes Energies, Universitat Autonoma de Barcelona, E-08193, Bellaterra (Barcelona), Spain} 
\affiliation{Baylor University, Waco, Texas  76798} 
\affiliation{Istituto Nazionale di Fisica Nucleare, University of Bologna, I-40127 Bologna, Italy} 
\affiliation{Brandeis University, Waltham, Massachusetts 02254} 
\affiliation{University of California, Davis, Davis, California  95616} 
\affiliation{University of California, Los Angeles, Los Angeles, California  90024} 
\affiliation{University of California, San Diego, La Jolla, California  92093} 
\affiliation{University of California, Santa Barbara, Santa Barbara, California 93106} 
\affiliation{Instituto de Fisica de Cantabria, CSIC-University of Cantabria, 39005 Santander, Spain} 
\affiliation{Carnegie Mellon University, Pittsburgh, PA  15213} 
\affiliation{Enrico Fermi Institute, University of Chicago, Chicago, Illinois 60637} 
\affiliation{Comenius University, 842 48 Bratislava, Slovakia; Institute of Experimental Physics, 040 01 Kosice, Slovakia} 
\affiliation{Joint Institute for Nuclear Research, RU-141980 Dubna, Russia} 
\affiliation{Duke University, Durham, North Carolina  27708} 
\affiliation{Fermi National Accelerator Laboratory, Batavia, Illinois 60510} 
\affiliation{University of Florida, Gainesville, Florida  32611} 
\affiliation{Laboratori Nazionali di Frascati, Istituto Nazionale di Fisica Nucleare, I-00044 Frascati, Italy} 
\affiliation{University of Geneva, CH-1211 Geneva 4, Switzerland} 
\affiliation{Glasgow University, Glasgow G12 8QQ, United Kingdom} 
\affiliation{Harvard University, Cambridge, Massachusetts 02138} 
\affiliation{Division of High Energy Physics, Department of Physics, University of Helsinki and Helsinki Institute of Physics, FIN-00014, Helsinki, Finland} 
\affiliation{University of Illinois, Urbana, Illinois 61801} 
\affiliation{The Johns Hopkins University, Baltimore, Maryland 21218} 
\affiliation{Institut f\"{u}r Experimentelle Kernphysik, Universit\"{a}t Karlsruhe, 76128 Karlsruhe, Germany} 
\affiliation{Center for High Energy Physics: Kyungpook National University, Daegu 702-701, Korea; Seoul National University, Seoul 151-742, Korea; Sungkyunkwan University, Suwon 440-746, Korea; Korea Institute of Science and Technology Information, Daejeon, 305-806, Korea; Chonnam National University, Gwangju, 500-757, Korea} 
\affiliation{Ernest Orlando Lawrence Berkeley National Laboratory, Berkeley, California 94720} 
\affiliation{University of Liverpool, Liverpool L69 7ZE, United Kingdom} 
\affiliation{University College London, London WC1E 6BT, United Kingdom} 
\affiliation{Centro de Investigaciones Energeticas Medioambientales y Tecnologicas, E-28040 Madrid, Spain} 
\affiliation{Massachusetts Institute of Technology, Cambridge, Massachusetts  02139} 
\affiliation{Institute of Particle Physics: McGill University, Montr\'{e}al, Canada H3A~2T8; and University of Toronto, Toronto, Canada M5S~1A7} 
\affiliation{University of Michigan, Ann Arbor, Michigan 48109} 
\affiliation{Michigan State University, East Lansing, Michigan  48824} 
\affiliation{University of New Mexico, Albuquerque, New Mexico 87131} 
\affiliation{Northwestern University, Evanston, Illinois  60208} 
\affiliation{The Ohio State University, Columbus, Ohio  43210} 
\affiliation{Okayama University, Okayama 700-8530, Japan} 
\affiliation{Osaka City University, Osaka 588, Japan} 
\affiliation{University of Oxford, Oxford OX1 3RH, United Kingdom} 
\affiliation{University of Padova, Istituto Nazionale di Fisica Nucleare, Sezione di Padova-Trento, I-35131 Padova, Italy} 
\affiliation{LPNHE, Universite Pierre et Marie Curie/IN2P3-CNRS, UMR7585, Paris, F-75252 France} 
\affiliation{University of Pennsylvania, Philadelphia, Pennsylvania 19104} 
\affiliation{Istituto Nazionale di Fisica Nucleare Pisa, Universities of Pisa, Siena and Scuola Normale Superiore, I-56127 Pisa, Italy} 
\affiliation{University of Pittsburgh, Pittsburgh, Pennsylvania 15260} 
\affiliation{Purdue University, West Lafayette, Indiana 47907} 
\affiliation{University of Rochester, Rochester, New York 14627} 
\affiliation{The Rockefeller University, New York, New York 10021} 
\affiliation{Istituto Nazionale di Fisica Nucleare, Sezione di Roma 1, University of Rome ``La Sapienza," I-00185 Roma, Italy} 
\affiliation{Rutgers University, Piscataway, New Jersey 08855} 
\affiliation{Texas A\&M University, College Station, Texas 77843} 
\affiliation{Istituto Nazionale di Fisica Nucleare, University of Trieste/\ Udine, Italy} 
\affiliation{University of Tsukuba, Tsukuba, Ibaraki 305, Japan} 
\affiliation{Tufts University, Medford, Massachusetts 02155} 
\affiliation{Waseda University, Tokyo 169, Japan} 
\affiliation{Wayne State University, Detroit, Michigan  48201} 
\affiliation{University of Wisconsin, Madison, Wisconsin 53706} 
\affiliation{Yale University, New Haven, Connecticut 06520} 
\author{T.~Aaltonen}
\affiliation{Division of High Energy Physics, Department of Physics, University of Helsinki and Helsinki Institute of Physics, FIN-00014, Helsinki, Finland}
\author{J.~Adelman}
\affiliation{Enrico Fermi Institute, University of Chicago, Chicago, Illinois 60637}
\author{T.~Akimoto}
\affiliation{University of Tsukuba, Tsukuba, Ibaraki 305, Japan}
\author{M.G.~Albrow}
\affiliation{Fermi National Accelerator Laboratory, Batavia, Illinois 60510}
\author{B.~\'{A}lvarez~Gonz\'{a}lez}
\affiliation{Instituto de Fisica de Cantabria, CSIC-University of Cantabria, 39005 Santander, Spain}
\author{S.~Amerio}
\affiliation{University of Padova, Istituto Nazionale di Fisica Nucleare, Sezione di Padova-Trento, I-35131 Padova, Italy}
\author{D.~Amidei}
\affiliation{University of Michigan, Ann Arbor, Michigan 48109}
\author{A.~Anastassov}
\affiliation{Rutgers University, Piscataway, New Jersey 08855}
\author{A.~Annovi}
\affiliation{Laboratori Nazionali di Frascati, Istituto Nazionale di Fisica Nucleare, I-00044 Frascati, Italy}
\author{J.~Antos}
\affiliation{Comenius University, 842 48 Bratislava, Slovakia; Institute of Experimental Physics, 040 01 Kosice, Slovakia}
% ...
\collaboration{CDF Collaboration\footnote{With visitors from $^a$University of Athens, 15784 Athens, Greece, 
$^b$Chinese Academy of Sciences, Beijing 100864, China, 
$^c$University of Bristol, Bristol BS8 1TL, United Kingdom, 
$^d$University Libre de Bruxelles, B-1050 Brussels, Belgium, 
$^e$University of California Irvine, Irvine, CA  92697, 
$^f$University of California Santa Cruz, Santa Cruz, CA  95064, 
$^g$Cornell University, Ithaca, NY  14853, 
$^h$University of Cyprus, Nicosia CY-1678, Cyprus, 
$^i$University College Dublin, Dublin 4, Ireland, 
$^j$University of Edinburgh, Edinburgh EH9 3JZ, United Kingdom, 
$^k$University of Heidelberg, D-69120 Heidelberg, Germany, 
$^l$Universidad Iberoamericana, Mexico D.F., Mexico, 
$^m$University of Manchester, Manchester M13 9PL, England, 
$^n$Nagasaki Institute of Applied Science, Nagasaki, Japan, 
$^o$University de Oviedo, E-33007 Oviedo, Spain, 
$^p$Queen Mary, University of London, London, E1 4NS, England, 
$^q$Texas Tech University, Lubbock, TX  79409, 
$^r$IFIC(CSIC-Universitat de Valencia), 46071 Valencia, Spain, 
}}
\noaffiliation

%% file: prl_wh1invfb.bbl
\begin{thebibliography}{20}
\bibitem[{coo()}]{coordinates}
\bibinfo{note}{The CDF reference frame uses cylindrical coordinates, where $z$
  is the direction of the proton beam, and $\theta$ and $\phi$ are the polar
  and azimuthal angles with respect to the proton beam. The pseudorapidity is
  defined as $\eta = -\ln \tan \theta/2$, the transverse energy $E_T=E\sin
  \theta$, and the transverse momentum $p_T=p \sin \theta$. The \met\ magnitude
  is the vector sum of all calorimeter deposits, projected into the transverse
  plane. The \met\ vector is corrected for the energy deposition of the
  high-energy muons as well as for the jet energy corrections.}
\end{thebibliography}
